\documentclass[prl,showpacs,twocolumn,preprintnumbers]{revtex4}
\usepackage{graphicx}
\usepackage{dcolumn}
\usepackage{amsmath}
\usepackage{bm}

\begin{document}

\title{Optical Vortices with Starlight: Implications for ground-based
stellar coronagraphy}

\author{F. Tamburini}

\author{G. Anzolin}

\author{G. Umbriaco}

\author{A. Bianchini}

\author{C. Barbieri}

\affiliation{Dipartimento di Astronomia, Universit\`{a} di Padova,
vicolo dell'Osservatorio 3, I-35122 Padova, Italy.}

\begin{abstract}
Using an $l = 1$ blazed fork-hologram at the focal plane of the
Asiago 122~cm telescope, we obtained optical vortices from the
stellar system Rasalgethi ($\alpha$~Herculis) and from the single
star Arcturus ($\alpha$~Bootis). We have analyzed the structure of
the optical vortices obtained from non-monochromatic starlight under
very poor seeing conditions using a fast CCD camera to obtain
speckle patterns and carry out the \textit{lucky imaging} technique,
alternative to adaptive optics. With the insertion of a red filter
and of a Lyot stop we performed $\ell = 1$ optical vortex
coronography the double star HD74010. The results are in agreement
with theory and numerical simulations. Our results open the way to
applications of optical vortices to ground based astronomical
observations, in particular for coronagraphy with $\ell > 1$ masks.
No intrinsic orbital angular momentum was detected in the starlight.
\end{abstract}


\maketitle

\textit{\textbf{Introduction} --} Optical vortices (OVs) are phase
defects embedded in light beams endowed with orbital angular
momentum (OAM). OVs, and more in general light beams carrying OAM,
are generated after the insertion of a phase modulating device (PMD)
that imprints a certain vorticity on the phase distribution of the
original beam. Such beams can be mathematically described by a
superposition of Laguerre-Gaussian (L-G) modes characterized by the
two integer-valued indices $\ell$ and $p$. The azimuthal index
$\ell$ describes the number of twists of the helical wavefront and
the radial index $p$ gives the number of radial nodes of the mode.
The electromagnetic field amplitude of a generic L-G mode in a plane
orthogonal to the direction of propagation is
\begin{equation} \label{eqn:lgmode}
 u_{\ell p}(r, \theta) \propto {\left(\frac{r \sqrt{2}}{w_0}\right)}^{|\ell|}
 L_p^{|\ell|} \left(\frac{2 r^2}{w_0^2}\right) \exp
 \left(-\frac{r^2}{w_0^2}\right) \exp (-\mathrm{i} \, \ell \, \theta)
\end{equation}
where $w_0$ is the beam radius and $L_n^m (x)$ is the associated
Laguerre polynomial. The presence of a phase factor $\exp
(-\mathrm{i} \, \ell \, \theta )$ implies that these cylindrically
symmetric modes carry an OAM equal to $\ell \, \hbar$ per photon,
relative to their symmetry axis~\cite{all92,vaz02}. For the same
reason, a phase singularity is embedded in the wavefront, all along
the propagation axis, with topological charge equal to $\ell$. The
intensity distribution of an L-G mode with $p = 0$ is generally
shaped as a ring with a central dark hole, where the intensity is
null due to total destructive interference. The radius of maximum
intensity of this \textit{donut} grows as the square root of $\ell$
and the intensity value decreases as $\ell^{-1/2}$~\cite{pad95}.

Experimentally, these properties are produced with beams propagating
through nonlinear optical systems~\cite{are91} and Kerr nonlinear
refractive media~\cite{swa92}. This induced to argue that OAM could
also be naturally generated by some astrophysical environments,
possibly related to turbulent interstellar media with density
discontinuities on wide scale ranges or to the distorted geometry
around Kerr black holes. It was suggested that OAM could be also
present in the blackbody radiation of the cosmic microwave
background~\cite{har03}. OVs have been already applied in diverse
research fields such as laboratory optics, nanotechnologies and
biology~\cite{gri03}. Two astronomical applications have been
indicated: first, to overcome the instrumental limitations in the
resolution of very close stellar sources due the Rayleigh
separability criterion in diffraction-limited
telescopes~\cite{swa01,palphd,tam06b}, second, to improve the
capability of imaging extrasolar planets by peering into the
darkness of an OV generated by a PMD inserted in the optical path of
a Lyot-type coronagraph~\cite{swa01,foo05,lee06}. Similar results
can be obtained by direct imaging and analysis of tiny deviations of
the main star's OV introduced by the presence of a close fainter
companion~\cite{tam06a,bar07}. However, ground-based telescopes will
always feel the detrimental effects of atmospheric turbulence
(seeing), which reduces the resolving power, even with adaptive
optics. Therefore, it is important for the above astronomical
applications of OVs to know how the seeing can affect the expected
\textit{donut} pattern generated by stellar sources. In this Letter
we present the first images of stellar OVs generated by a PMD placed
in the optical path of the Asiago 122~cm telescope. In particular,
we show how the effects of the atmospheric turbulence can be
overcome to produce good quality OVs suitable for OV coronagraphy
even under poor seeing conditions and in white light. Even if the
contrast needed for the search of exoplanet or for classical
coronagraphy can be achieved mainly with even $\ell$s~\cite{maw05},
we decided to use $\ell = 1$ OVs to avoid some experimental
complications as higher order ones may not remain stable while
propagating in turbulent media, and may split in first order
ones~\cite{nye74}.

\textit{\textbf{Non-monochromatic optical vortices} --} Several
types of PMDs have been designed to generate axially symmetric OVs
from an incident monochromatic on-axis beam. The most efficient are
computer-generated fork-holograms~\cite{baz90} and spiral phase
plates (SPP)~\cite{bei94}. However, the use of non-monochromatic
beams in astronomical applications of OVs is mandatory to collect
enough photons from faint stellar objects. In this case, the
\textit{donut}-shaped structure of monochromatic OVs will be
changed~\cite{shv05,pal04}. Using SPPs, which are helicoidal
transmission optical devices with a given total thickness variation
$h_s$, all monochromatic OVs will have the same axis of symmetry.
However, different wavelengths will present different OAM values
according to $\ell = \Delta n(\lambda) \, h_s / \lambda$, where
$\Delta n$ is the difference between the refraction indices of the
SPP material and the surrounding medium. Thus, the transmitted beam
will possess different OAM values producing \textit{donut}-shaped
patterns of different sizes and a progressive filling of the central
dark region. Recently~\cite{swa06} a scheme has been proposed to
overcome this problem, but achromaticity is expected only for a
limited bandwidth ($\sim 100$~nm) in the visible. We have instead
used a fork-hologram which is a grating with a number $l$ of
dislocations on its center. Monochromatic on-axis beams produce OVs
with OAM indices $\ell = m \, l$, where $m$ is the diffraction order
of the grating. Differently from SPPs, fork-holograms present the
advantage of generating OVs with the same $\ell$ at all wavelengths
for on-axis polychromatic sources. For off-axis sources $\ell$
decreases as the star moves away from the center of the hologram.

Spatial coherence of the incident beam is generally assumed to hold
only for stellar sources when observed from space, but when
ground-based telescopes are used, quite dramatic effects are
introduced by the atmospheric turbulence and spatial coherence is
lost. However, for exposures shorter than the turbulence timescale
($\sim 10-100$~ms in the optical/near-infrared) we can obtain a
group of bright speckles that represent the interference image
produced by the coherent wavefronts generated by the random
distribution of the atmospheric irregularities. A single nearly
diffraction-limited stellar image can occasionally be produced when
most of the stellar light falls in a single bright speckle.
Fried~\cite{fri78} coined the term \textit{lucky exposures} to
describe high quality short exposures occurring in such a fortuitous
way. This is the basis of \textit{lucky imaging}~\cite{law06}, one
of the speckle imaging techniques used in modern astronomy as
alternative to adaptive optics. Here we use the \textit{lucky
imaging} technique to obtain OVs from stellar sources useful for OV
coronagraphy, with the additional restriction of selecting only
those exposures where the star to be fainted is centered with the
hologram dislocation.

\begin{figure}
\centering
\includegraphics[width=6cm, keepaspectratio]{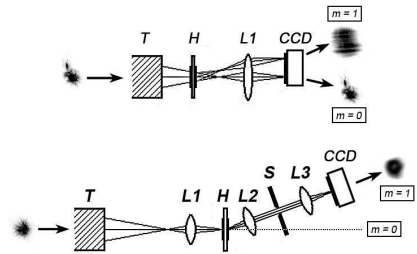}
\caption{Optical setups, without (upper panel) and with (lower
panel) spatial filter. \textit{T} is the telescope; \textit{L1, L2,
L3} the lenses; \textit{H} the $l = 1$ fork-hologram; \textit{S} the
spatial filter slit. Stellar speckle patterns are sketched on the
left of the optical setups, while the output images at the 0th and
1st diffraction order, on the right. All angles and displacements
are exaggerated for clarity.}
\label{fig:telescope}
\end{figure}

\textit{\textbf{Telescope Observations} --} Here we report the
results obtained in a campaign of observations carried out from 2005
to 2007 with the Asiago 122~cm telescope, that followed some
preliminary laboratory simulations~\cite{tam06a}. We first observed
the multiple system Rasalgethi ($\alpha$~Her) and the single star
Arcturus ($\alpha$~Boo). $\alpha$~Her is a visual binary composed by
two unresolved binary systems presently separated by $4''.7$:
$\alpha$~Her~A, formed by an M5~Ib-II semiregular variable ($m_V =
2.7 - 4.0$) and a fainter companion separated by
$0''.19$~\cite{mca89}, and $\alpha$~Her~B, containing a G0~II-III
giant ($m_V = 5.4$) and a fainter secondary separated by
$0''.0035$~\cite{hal81}. $\alpha$~Boo, instead, is a single star
having visual magnitude $m_V = 0.04$ and spectral type K1.5~III. We
also performed $\ell = 1$ OV coronography with the double system
HD74010, which F0 components are separated by $9''.9$ and have
visual magnitudes $m_{V,A} = 7.72$ and $m_{V,B} = 7.87$. The optical
schemes for the two types of observations to image the OVs are
sketched in Fig.~\ref{fig:telescope}; the OV coronagraph is
presented in~\cite{lee06}. Our $l = 1$ fork-hologram \textit{H} is
blazed at the first diffraction order with 20 lines $\text{mm}^{-1}$
and has an active area of $2.6 \times 2.6 \text{ mm}^2$. \textit{H}
was placed in proximity of the F/16 Cassegrain focus of the
telescope. The OVs produced by non-monochromatic light beams
crossing the fork-hologram show intensity patterns that appear as
rings stretched along the dispersion direction with a
\textit{central dark strip}. The spectral dispersion also causes a
partial filling of the central dark zone. Thus, if we want to use
non-monochromatic OVs produced by fork-holograms for OV
coronagraphy, we must limit the spectral range~\cite{tam06a} and/or
restore the \textit{donut} shape (see e.~g.~\cite{lea03}). In our
case we used a variable spatial filter, \textit{S}, made by a slit
placed on the Fourier plane of the collimating lens \textit{L2} to
limit the dispersion of the light at the first diffraction order.
This adjustable slit works as a tunable bandpass filter with flat
spectral response and has been used only for single on-axis stars.
We used a fast CCD camera with $660 \times 494$ and $7.4 \times
7.4$~$\mu$m pixels, 16 bit dynamical range (4000 to 6700~\AA\ with a
peak at 5200~\AA) to image the OVs' shapes, and, for coronography,
the ultrafast EMCCD Andor iXon+ camera.

\begin{figure}
\centering
\includegraphics[width=5cm, keepaspectratio]{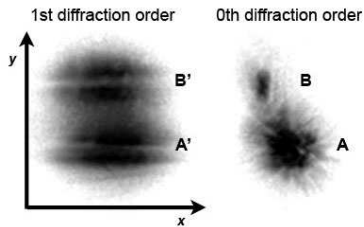}
\caption{Speckle pattern (right) and dispersed OVs (left) obtained
by summing the 4 \textit{lucky} exposures of $\alpha$~Herculis. The
image is displayed in a squared greyscale. The reference system used
for the extraction of the OV profiles is reported. The $x$ axis
corresponds to the direction of dispersion.}
\label{fig:rasal}
\end{figure}

\begin{figure}
\centering
\includegraphics[width=7.5cm, keepaspectratio]{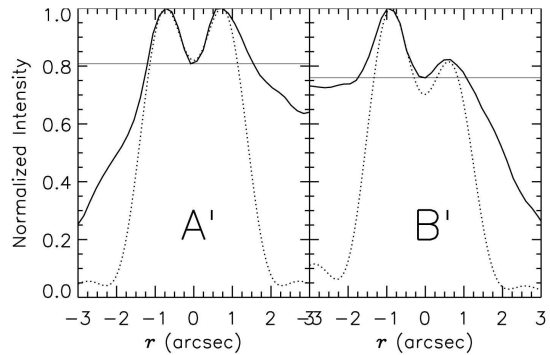}
\caption{Observed OV profiles of $\alpha$~Her~A ($A'$, left) and $\alpha$~Her~B
($B'$, right) obtained from the selected \textit{lucky} exposures.
The corresponding numerical simulations are represented by the dotted lines.
The thin lines are drown at any dip level.}
\label{fig:profiles}
\end{figure}

While observing the multiple system $\alpha$~Her we could
simultaneously observe the zeroth and the first diffraction order
($\ell = 0$ and $\ell = 1$, respectively, for on-axis stars). We can
then verify how the observed OV patterns change with the distance
of the beam from the center of the hologram. We placed the
fork-hologram 30~mm before the telescope focal plane and set the
main star ($\alpha$~Her~A) at the center of the optical system. The
mean seeing was $\sim 3''.8$. Under such bad seeing conditions the
\textit{lucky imaging} technique empirically predicts that the
resolution should improve by a factor 2 if we select 1\% of frames
taken with an exposure time of $\sim 80$~ms~\cite{law06}. We then
recorded a sequence of 860 frames at a time step of 70~ms. To select
the best exposures we imposed the additional restriction that the
intensity of the peaks of the $\alpha$~Her~A OV were equal, that
corresponded to the stellar beam intersecting the center of the
hologram. Fig.~\ref{fig:rasal} represents the sum of the 10\%
\textit{lucky} speckle patterns observed together with their
corresponding OVs. Since the signal from $\alpha$~Her~A is
saturated, an angular resolution of $\sim 1''.5$ was derived from
the full width at half maximum (FWHM) of the intensity profile of
$\alpha$~Her~B. Being the central part of the OV profile of
$\alpha$~Her~A saturated, the choice of the \textit{lucky} exposures
has been made by analyzing the intensity profiles taken across the
wings of the dispersed OVs. In Fig.~\ref{fig:profiles} we analyze in
detail the OV profiles averaged over 10 pixels wide strips
perpendicular to the dispersion, i.~e. along the $y$ axis of
Fig.~\ref{fig:rasal}. We find that the ratio of the two maxima
fitted with two Gaussians is $R_{A'} = 0.995 \pm 0.005$ for the
on-axis star and $R_{B'} = 0.83 \pm 0.01$ for the off-axis star.
We also numerically reconstructed the theoretical
intensity profiles of the chromatically dispersed on-axis and off-axis
OVs. These profiles present dips with intensity of 82\% and 70\%
respectively. As expected for $\ell = 1$, their central regions can
never be totally obscured also under ideal seeing conditions. The
profile $A'$ of $\alpha$~Her~A shows a central dip of 81\%.
The OV generated by $\alpha$~Her~$B$, being
off-axis, presents an asymmetric shape with a non-integer $\ell$
value.

\begin{figure}
\centering
\includegraphics[width=6cm, keepaspectratio]{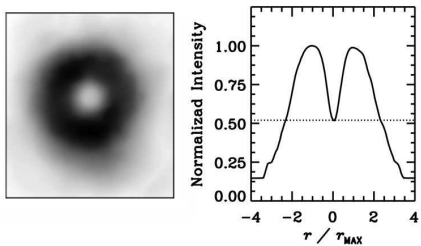}
\caption{Left panel: the OV obtained by summing the selected 2\%
good frames (see text). Left panel: profile of the OV across the
direction perpendicular to the dispersion.}
\label{fig:arcturus}
\end{figure}

The single star $\alpha$~Boo was set at the center of the hologram
and the spatial filter \textit{S} was introduced to produce a
nearly-monochromatic circularly symmetric OV. We adopted a slit
width of 0.1~mm, corresponding to a 300~\AA\ bandpass width in the
visible, that ensured enough S/N ratio for the 70~ms exposures.
Since in this case the slit did not allow the simultaneous
observation of the speckle patterns at the zeroth diffraction order,
we could only select a 17\% sub-sample of frames that presented OVs
circularly symmetric and therefore centered with the optical
singularity of the hologram. We assumed that this condition was met
when the ratios of the two intensity peaks measured along two
perpendicular axes across the OV were close to 1 within an error of
$\sim 10^{-2}$. Fig.~\ref{fig:arcturus} shows the OV obtained by
summing the selected \textit{lucky} frames. The central region of
the OV is not totally dark because of the loss of the starlight
coherence due to the very bad seeing conditions and to the presence
of residual chromatism. We however notice that the contrast between
the dark center and the bright ring improves down to 52\% with
respect to the previous unfiltered symmetric OVs.

\begin{figure}
\centering
\includegraphics[width=6cm, keepaspectratio]{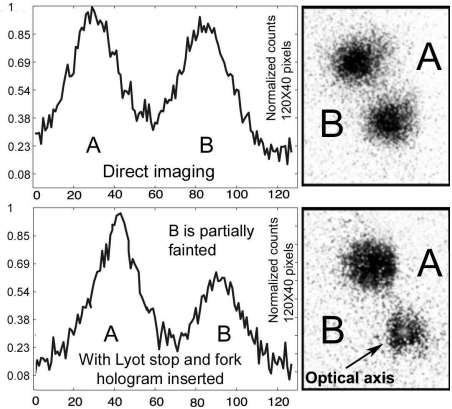}
\caption{$\ell = 1$ OV coronagraphy of HD74010. Left panel: the
averaged intensity profiles of the double star before (up) and after
(down) the insertion of the Lyot stop. Right panel: the
corresponding snapshots (see text).}
\label{fig:ovc}
\end{figure}

We finally observed the double system HD74010 through a 100~\AA\
bandpass red filter centered at 6532~\AA\ without and with a Lyot
stop. The exposure times were 0.01~s. Fig.~\ref{fig:ovc} shows the
results of $\ell = 1$ OV coronagraphy: the intensity profiles of the
two stars without and with the Lyot mask, obtained by averaging over
a 40 pixels strip, are shown in the left panels. Right panels show
the corresponding snapshots. The on-axis component B of the binary
system appears fainted by a factor $\sim 1.7$, close to the factor 2
derived from numerical simulations. The partial obscuration of the
on-axis star is due to the fact that we were using a $\ell = 1$
fork-hologram. Total obscuration may be mainly achieved with even
$\ell$ values.

\textit{\textbf{Discussion and Conclusions} --} We demonstrated the
feasibility of OV coronagraphy even under very bad seeing conditions
with ground-based telescopes. Using the \textit{lucky imaging}
technique to improve the quality of our images, we obtained slightly
symmetric OV patterns from on-axis starlight beams and performed a
first trial of OV coronagraphy obtaining a fainting close to that
that can be actually obtained with our $l = 1$ fork-hologram. We
tested how the pattern of polychromatic starlight OVs depends from
the off-axis position of the star beam. We showed how
polychromaticity and the lack of coherence produced by the
atmospheric turbulence may heavily alter the theoretical
\textit{donut} profile of the OV reducing the contrast needed for
coronagraphy. Although only even values of $\ell$ can produce
perfect rejection of the on-axis starlight as required for
coronagraphy, we decided to start with an $l = 1$ fork-hologram to
ensure the stability of the OVs along the optical path. This
explains the importance of this choice as a first test of OV
coronagraphy. We found that our experimental results are consistent
with theoretical predictions. We suggest that these new promising
techniques could find their best application mainly at telescopes
with adaptive optics, or in space instruments. Finally, since we did
not detect any OV at the other diffraction orders of the hologram,
we can conclude that starlight does not have an appreciable
intrinsic OAM.

\acknowledgments
We would like to thank the Institut f\"{u}r
Experimentalphysik, University of Wien, Zeilinger-Gruppe for
support, helpful discussions and comments. This work has been partly
supported by the University of Padova.

\end{document}